\documentstyle[12pt,epsfig]{article}

\input epsf

\newcommand{\be}{\begin{equation}}
\newcommand{\ee}{\end{equation}}
\newcommand{\ba}{\begin{array}}
\newcommand{\ea}{\end{array}}
\newcommand{\baa}{\begin{eqnarray}}
\newcommand{\btab}{\begin{tabular}}
\newcommand{\etab}{\end{tabular}}
\newcommand{\eaa}{\end{eqnarray}}

\newcommand{\lab}[1]{\label{#1}}
\newcommand{\re}[1]{\ref{#1}}

\newcommand{\mb}[1]{\mbox{#1}}
\newcommand{\ul}[1]{\underline{#1}}
\newcommand{\edd}{\end{document}}

\newcommand{\alf}{\ifmmode\alpha \else$\alpha \ $\fi}
\newcommand{\bt}{\ifmmode\beta \else$\beta \ $\fi}
\newcommand{\gm}{\ifmmode\gamma \else$\gamma \ $\fi}
\newcommand{\Dl}{\ifmmode\Delta \else$\Delta \ $\fi}
\newcommand{\eps}{\ifmmode\varepsilon \else$\varepsilon \ $\fi}
\newcommand{\dl}{\ifmmode\delta \else$\delta \ $\fi}
\newcommand{\et}{\ifmmode\eta \else$\eta \ $\fi}
\newcommand{\vph}{\ifmmode\varphi \else$\varphi \ $\fi}
\newcommand{\om}{\ifmmode\omega \else$\omega \ $\fi}
\newcommand{\pl}{\ifmmode\partial \else$\partial \ $\fi}
\newcommand{\ps}{\ifmmode\psi \else$\psi \ $\fi}
\newcommand{\sg}{\ifmmode\sigma \else$\sigma \ $\fi}
\newcommand{\phf}{\ifmmode\varphi^4 \else$\varphi^4 \ $\fi}
\newcommand{\Lam}{\ifmmode\Lambda \else$\Lambda$\fi}
\newcommand{\frc}[2]{{#1/#2}}

\newcommand{\ppp}[1]{%
        \setbox0=\hbox{#1}%
        \kern-.02em\copy0\kern-\wd0
        \kern+.04em\copy0\kern-\wd0
        \kern-.02em\raise.0217em\box0}

\newcommand{\lsim}{
 \mathrel{\setbox0=\hbox{$<$}\raise0.6ex\copy0\kern-\wd0
 \lower0.65ex\hbox{$\sim$}}}

\newcommand{\gsim}{
 \mathrel{\setbox0=\hbox{$>$}\raise0.6ex\copy0\kern-\wd0
 \lower0.65ex\hbox{$\sim$}}}

\newcommand{\etal}{{\em et al.}}

\newcommand{\PRD}[3]{Phys.\ Rev.\ D {\bf {#1}}, {#2} ({#3})}

\newcommand{\PRL}[3]{Phys.\ Rev.\ Lett.\ {\bf {#1}}, {#2} ({#3})}
\newcommand{\NPA}[3]{Nucl.\ Phys.\ A {\bf {#1}}, {#2} ({#3})}
\newcommand{\NPB}[3]{Nucl.\ Phys.\ B {\bf {#1}}, {#2} ({#3})}
\newcommand{\PLB}[3]{Phys.\ Lett.\ B {\bf {#1}}, {#2} ({#3})}
\newcommand{\ZPC}[3]{Z. Phys.\ C {\bf {#1}}, {#2} ({#3})}
\newcommand{\JPG}[3]{J. Phys.\ G {\bf {#1}}, {#2} ({#3})}
\newcommand{\EJP}[3]{Eur. J. Phys.\ C {\bf {#1}}, {#2} ({#3})}
\newcommand{\EJPA}[3]{Eur. J. Phys.\ A {\bf {#1}}, {#2} ({#3})}

\begin{document}
%

\begin{titlepage}
\renewcommand{\thefootnote}{\fnsymbol{footnote}}
\makebox[2cm]{}\\[-1in]
\begin{flushright}
\begin{tabular}{l}
TUM/T39-99-04\\
hep-ph/9903???
\end{tabular}
\end{flushright}
\vskip0.4cm
\begin{center}
  {\Large\bf
Conformal string operators and evolution of
skewed parton distributions}\footnote{work supported in part by BMBF}

\vspace{2cm}

N. Kivel\footnote{Alexander von Humboldt fellow}$^{a b}$, L. Mankiewicz$^{a c}$

\vspace{1.5cm}

\begin{center}

{\em$^a$Physik Department, Technische Universit\"{a}t M\"{u}nchen, 
D-85747 Garching, Germany} 

{\em $^b$Petersburg Nuclear Physics Institute,
  188350, Gatchina, Russia
}

{\em $^c$N. Copernicus Astronomical Center, ul. Bartycka 18,
PL--00-716 Warsaw, Poland}

\end{center}

\vspace{1cm}


\vspace{3cm}

\centerline{\bf Abstract}
\begin{center}
\begin{minipage}{15cm}

We have investigated skewed parton distributions in coordinate
space. We found that their evolution can be described in a simple manner in
terms of non-local, conformal operators introduced by
Balitsky and Braun. The resulting formula is given by a Neumann series
expansion. Its structure
resembles, for all values of the asymmetry parameter, the well-known solution
of the 
ERBL equation in the momentum space. Performing Fourier transformation  we have
reproduced known results for evolution of momentum-space distributions.

\end{minipage}
\end{center}

\end{center}
\end{titlepage}
\setcounter{footnote}{0}

\newpage

\section{Introduction}

Recently, a lot of effort has been invested in exploring deeply-virtual
Compton scattering and hard, exclusive meson production processes, see
\cite{Ji98a,Guich98} 
and \cite{HERA,HERMES,COMPASS,TJNAF} for recent reviews of theoretical and
experimental situation. 
The celebrated factorization theorems \cite{CFS97,Rad97,Ji98b} assure that
when photon virtuality is large the underlying photon-parton sub-processes are
indeed dominated by  
short distances and can be calculated perturbatively. The necessary
information about long-distance, non-perturbative dynamics enters in the form
of 
distribution amplitudes of the produces mesons, and new, generalized
\cite{Ji97,Rad97} parton distribution functions of the hadronic target.
Factorizable hard, exclusive processes provide therefore a possibility to study
new aspects of nucleon structure, which cannot be accessed in inclusive
measurements. 

More detailed studies of hard, exclusive meson production
\cite{Hood98,MPW98a,Vanderh98,MPW98b,MPR99,FPPS99} have
demonstrated
that 
cross-section of these processes should be sufficiently large, so they
can be investigated in
the next round of dedicated experiments. 
These estimates are based on leading order (LO) QCD calculations which
typically neglect  the scale-dependence of 
generalized parton distributions, except for a diffractive region
\cite{Rys93,FS94}. There, one can 
argue \cite{MPW98a,Rad98c} that generalized parton distributions which
determine the 
dominant, 
imaginary 
part of the amplitude are proportional to the corresponding forward parton
distributions. On the other hand, kinematical conditions for future
measurements are such that
a lot of data will be collected outside the
diffractive region.
Once data points achieve sufficient accuracy, a practical discussion of
scale dependence of amplitudes of hard, exclusive processes will become
unavoidable. Evolution equations for generalized parton distributions can be
found in \cite{BalBra88,Evolop,Ji97}. Their general properties have been
discussed at length in the literature \cite{Ji97,Rad97,Rad99}, and by now
various 
numerical algorithms \cite{FFGS97,Bel97b,MPW98a,Bel98a,AMar99} and
theoretical methods \cite{Shuv99} have been proposed and tested 
numerically. Most of these methods make more or less explicit use of the
local conformal operators which are multiplicatively
renormalized 
at the one-loop level \cite{ER78,Conformal}. Note 
also that so far the main effort has been devoted to
study evolution of skewed parton distributions themselves, while away from the
diffractive region the real issue is the scale dependence of the QCD amplitude.

The goal of this paper is 
twofold. First, we aim at a new,
practically applicable method
of treating the problem of 
scale-dependence of skewed parton distributions and corresponding
amplitudes of hard, exclusive processes. Second, we want to explore the
explicit operator 
solution to one-loop QCD evolution equations found by Balitsky and Braun, which
involves non-local conformal operators. Although it
has been found about a decade ago, 
we are not aware about any previous
attempt to make phenomenological use of it.  As the
Balitsky-Braun solution was formulated in
terms of coordinate-space quark string operators it is natural to
introduce here a concept of coordinate-space skewed parton distributions which
are generalization of coordinate-space 
forward parton distributions \cite{Iofdst}. As we shall
argue below,
hard exclusive 
processes involve momentum- and coordinate-space parton distributions in a
rather 
symmetric way, so both languages are equally convenient from
the point of 
view of practical calculations. 

It turns out that the language of coordinate-space string operators can
indeed be applied successfully 
to the problem of finding a solution of the evolution equations for
coordinate-space 
distributions. The result can be interpreted as a Neumann series
expansion of the latter in terms of matrix elements of multiplicatively
renormalizable, {\em non-local} conformal operators. It can be understood also
as an
operator identity - an expansion of QCD string operator in terms of
non-local conformal operators.

Having found a solution of the evolution equations in coordinate-space, we
have investigated also a corresponding scheme for evolution of momentum-space
distributions. 
Here, we have
found that coordinate-space considerations suggest splitting of initial
conditions into two pieces, corresponding to initial parton
distributions with average momentum larger and smaller than the asymmetry
parameter. Because of the linear character of the evolution equations, both
terms 
evolve independently. The resulting algorithm is a combination of expansion in
terms of local conformal operators of one part, and general expansion in terms
of orthogonal polynomials of the other. Both techniques have been already
discussed in the literature \cite{Bel97a,Bel97b,MPW98a,Bel98a}. Here we 
contribute to this 
discussion by 
showing
how the corresponding formulae obey general self-consistency constraints.

Our presentation is organized as follows. 
In the following section we introduce a basic
framework for coordinate-space skewed parton distributions. 
In the next section we present a
solution to the problem of scale dependence of the coordinate-space
distributions as an expansion in terms of Bessel functions. 
Armed with this result, we show that it can be
interpreted as 
an expansion in terms of matrix elements of non-local, conformal operators and
deduce the corresponding operator identity.
Our result is closely related to the 
explicit operator
solution found in \cite{BalBra88}. The equivalence of both approaches is
explicitely demonstrated in the Appendix A.

In the next section we turn our attention to momentum-space
quark distributions. We will argue that coordinate-space considerations suggest
a 
particular algorithm for treating evolution in the momentum space, and discuss
general constraints which any such solution has to fulfill. The next section
contains 
a summary. Finally, two appendices are devoted to mathematical details which
are not discussed in the main body of the paper.

\section{Coordinate-Space Skewed Parton Distributions}

In this section, we introduce the basic theoretical description of
coordinate-space skewed parton distributions. Consider the twist-2, gauge
invariant, light-cone quark string operator normalized at a scale $\mu^2$:
\be\lab{eq:Q_op_def}
O(\alpha,\beta) = {\bar q}(\frac{\alpha+\beta}{2}z) \hat{z}
\left[\frac{\alpha+\beta}{2}z,\frac{\alpha-\beta}{2}z\right]
q(\frac{\alpha-\beta}{2}z)_{z^2 = 0}\, .
\ee
Here,$[a,b]$ denotes the 
path-ordered exponential  
$$
[a,b] = \,
{\cal P} \exp [ -i g 
\-\int_b^a  A^{\mu}(x) \, d x_\mu]
$$  
which reduces to $1$
in the Schwinger gauge $z\cdot A=0$ ($g$ stands for the strong coupling
constant and $A^{\mu}$ denotes the gluon field). Coordinates $\alpha$ and
$\beta$ are defined in such a way that $\alpha\, z$ defines the center
of the string composed from quark fields and the gluon line between them while
$\beta\, z$ corresponds to its 'length', understood here simply as the
difference between  coordinates of the quark operators.

Charge conjugation odd and even quark string operators are obtained from
$O(\alpha,\beta)$ by taking its  components symmetric, respectively
antisymmetric in $\beta$. Denoting the former and the latter by
$O^+(\alpha,\beta)$, respectively $Q^-(\alpha,\beta)$ one has therefore
\be\lab{eq:O_C_decomp}
O^\pm(\alpha,\beta) = \frac{1}{2}\left(O(\alpha,\beta) \pm O(\alpha,-\beta)
\right) \, . 
\ee

Factorizable hard exclusive processes are described by matrix elements of
$O(\alpha,\beta)$ between nucleon states $\left \langle n(P',S')\right|$ and
$\left| n(P,S)\right \rangle$ with the corresponding four-momenta  $P$, $P'$  
and spins $S$, $S'$. Here, the momentum space skewed parton
distributions enter naturally \cite{Ji97,Rad97}. In the following, we have
chosen to use 
convention introduced by Ji \cite{Ji97}. In this notation, the matrix element
(\re{eq:Q_op_def}) can be represented as:
\begin{eqnarray}\label{def:corr_F_xi}
&&\hspace*{-1cm} 
\left \langle n(P',S')\right| 
O(\alpha,\beta) 
\left|n(P,S)\right \rangle
= 
\nonumber \\
&&
\bar N(P',S')\,\hat z\, N(P,S) e^{-i \alpha \frac{r}{2}\cdot z}
\!\int_{-1}^1 \!\! du \,F(u,\xi;\mu^2) e^{i u\beta({\bar P}\cdot z)}
 + 
\nonumber \\
&&\hspace{-1cm} 
\bar N(P',S')\frac{\sigma_{\mu\nu}z^{\mu}r^{\nu}}{i M}  N(P,S)
e^{-i \alpha \frac{r}{2}\cdot z}
\!\int_{-1}^1 \!\! du \,K(u,\xi;\mu^2) e^{i u\beta({\bar P}\cdot z)} \, .
\end{eqnarray}
Here $N(P,S)$ and $\bar N(P',S')$ denote initial and final nucleon spinors,
respectively. The average nucleon momentum is denoted by $\bar P = (P +
P')/2$, and the momentum transfer is $r = P-P'$. The asymmetry parameter $\xi$
is defined by $r\cdot z = 2 \xi {\bar P}\cdot z$.

From now on, we will
explicitly consider the skewed quark distribution $F(u,\xi;\mu^2)$ only.
Evolution equations for $K(u,\xi;\mu^2)$ are exactly the same, so one can
replace $F(u,\xi;\mu^2)$ by $K(u,\xi;\mu^2)$ at any stage of the following
presentation. To proceed further, we separate the symmetric and antisymmetric
part of $F(u,\xi;\mu^2)$, such that
\baa\lab{eq:Mat_el_split}
&&\!\int_{-1}^1 \!\! du \,F(u,\xi;\mu^2) e^{i u\beta({\bar P}\cdot z)} =
\nonumber \\
&&\!\int_{0}^1 \!\! du \,F^S(u,\xi;\mu^2) \cos{\left[u\beta({\bar P}\cdot
z)\right]} + i\, \!\int_{0}^1 \!\! du \,F^A(u,\xi;\mu^2)
\sin{\left[u\beta({\bar P}\cdot z)\right]}
\eaa
Obviously, matrix elements of $O^+(\alpha,\beta)$ are parametrized by
$F^S(u,\xi;\mu^2)$, or 'valence' (quark minus antiquark) skewed quark
distributions, while matrix elements of $O^-(\alpha,\beta)$ are given by
$F^A(u,\xi;\mu^2)$, or 'quark plus antiquark' distributions.

Note that due to charge conjugation and time-reversal invariance,
$F^S(u,\xi;\mu^2)$ and 
$F^A(u,\xi;\mu^2)$ contribute to the    
corresponding amplitudes $M^S(\xi;\mu^2)$ and $M^A(\xi;\mu^2)$ of hard,
exclusive processes 
through
\baa\lab{eq:Amplitudes} 
M^S(\xi;\mu^2) &\propto&
\!\int_{0}^{1} \!\! du \,F^S(u,\xi;\mu^2)\, \left[\frac{1}{\xi-u-i\epsilon} +
\frac{1}{\xi+u-i\epsilon}\right]
\nonumber \\
M^A(\xi;\mu^2) &\propto&
\!\int_{0}^{1} \!\! du \,F^A(u,\xi;\mu^2)\, \left[\frac{1}{\xi-u-i\epsilon} -
\frac{1}{\xi+u-i\epsilon}\right] \, .
\eaa

Now we introduce, in analogy to the forward case,  {\it coordinate-space}
skewed quark distributions ${\cal 
F}^S(\beta,\xi;\mu^2)$ and ${\cal F}^A(\beta,\xi;\mu^2)$ as 
\baa\lab{eq:MSD_to_CSD}
{\cal F}^S(\beta,\xi;\mu^2) &=& \frac{2}{\pi}\!\int_{0}^1 \!\! d\omega
\,F^S(\omega,\xi;\mu^2) \, \cos{\left(\omega \beta \right)} \, ,
\nonumber \\ 
{\cal F}^A(\beta,\xi;\mu^2) &=& \frac{2}{\pi}\!\int_{0}^1 \!\! d\omega
\,F^A(\omega,\xi;\mu^2) \, \sin{\left(\omega \beta \right)} \, .
\eaa 
The inverse transformation reads
\baa\lab{eq:CSD_to_MSD}
F^S(u,\xi;\mu^2) &=& \!\int_{0}^\infty \!\! d\beta
\,{\cal F}^S(\beta,\xi;\mu^2) \, \cos{\left(u \beta \right)} \, ,
\nonumber \\
F^A(u,\xi;\mu^2) &=& \!\int_{0}^\infty \!\! d\beta
\,{\cal F}^A(\beta,\xi;\mu^2) \, \sin{\left(u \beta \right)} \, .
\eaa 
Inserting these formulae into (\ref{eq:Amplitudes}) one finds that the
amplitudes 
$M^S(\xi;\mu^2)$ and $M^S(\xi;\mu^2)$ can be calculated 
directly in terms of coordinate-space skewed
quark distributions as
\baa \lab{eq:M_CSD_def}
M^S(\xi;\mu^2) &\propto& i \pi \!\int_{0}^\infty \!\! d\beta \, e^{-i \beta
\xi} \, {\cal F}^S(\beta,\xi;\mu^2) 
\nonumber \\ 
M^A(\xi;\mu^2) &\propto& i \pi \!\int_{0}^\infty \!\! d\beta \, e^{-i \beta
\xi} \, {\cal F}^A(\beta,\xi;\mu^2) \, . 
\eaa

Note that everywhere above we have used the notation $\beta$ for a variable
which arises from the string length $\beta$, introduced in (\ref{eq:Q_op_def}),
by a rescaling $\beta \rightarrow \beta ({\bar P}\cdot z)$.

Equations (\ref{eq:Amplitudes}) and (\ref{eq:M_CSD_def}) show that in general
momentum-
and coordinate-space distributions play a symmetric role the in description of
hard, exclusive processes. This is a new situation; in the case
of DIS only the former have a direct physical interpretation in terms of
observables. 
In particular, equations (\ref{eq:M_CSD_def}) can be
reinterpreted as  
\baa
M^S(\xi;\mu^2) &\propto& \left \langle n(P',S')\right| 
\!\int_{0}^\infty \!\! d\beta \, e^{-i \beta\xi {\bar P}\cdot z} \,
O^+(0,\beta)
\left| n(P,S)\right \rangle
\nonumber \\ 
M^A(\xi;\mu^2) &\propto& \left \langle n(P',S')\right|
\!\int_{0}^\infty \!\! d\beta \, e^{-i \beta \xi{\bar P}\cdot z} \, 
O^-(0,\beta)
\left| n(P,S)\right \rangle
\eaa
We have found it interesting to note that it is as convenient
to
discuss 
hard, exclusive processes in terms of the correlation function 
(\ref{def:corr_F_xi}) at various longitudinal distances, as it is in terms of
momentum-space parton distribution functions.

For numerical calculations it is often more convenient to rewrite the
amplitude $M^A(\xi;\mu^2)$ as
\be\lab{eq:Amp_a_1}
M^A(\xi;\mu^2) \propto \frac{1}{\xi}
\!\int_{0}^{1} \!\! du \, u \,F^A(u,\xi;\mu^2)\,
\left[\frac{1}{\xi-u-i\epsilon} + \frac{1}{\xi+u-i\epsilon}\right] \, .
\ee
The corresponding expression in terms of coordinate-space distribution reads:
\be
M(\xi;\mu^2)^A \propto i \pi \!\int_{0}^\infty \!\! d\beta \, e^{-i \beta \xi}
\, {\cal F}^{A\,\prime}(\beta,\xi;\mu^2) \, , 
\ee
where we have introduced a special notation
$$
{\cal F}^{A\,\prime}(\beta,\xi;\mu^2) \equiv  \frac{1}{\xi} \frac{d}{d\, \beta}
{\cal F}^{A}(\beta,\xi;\mu^2) \, .
$$

It is easy to see from (\re{eq:MSD_to_CSD}) and (\re{eq:CSD_to_MSD}) that
coordinate-space skewed distributions introduced here are generalizations of
coordinate-space parton distributions known from the forward case. On the
other hand, the physical interpretation of the coordinate-space skewed quark
parton distributions is twofold. First, they describe, according to
(\ref{def:corr_F_xi}), non-forward matrix elements of the non-local string
operator $Q(\alpha,\beta)$. Second, through (\re{eq:M_CSD_def}), they directly
determine amplitudes $M^S(\xi;\mu^2)$ and $M^A(\xi;\mu^2)$.  Comparing with
the forward case one finds that the way nucleon structure can be probed in
hard exclusive processes is indeed complementary to what happens in
deep-inelastic scattering experiments, as the latter are sensitive only to the
imaginary part of the amplitude $M^A$\footnote{In the forward, DIS case, the
  skewed quark distributions become the usual quark densities i.e.,
  $F^A(u,\xi;\mu^2) \rightarrow F^A(u,0;\mu^2) = q(u;\mu^2)$}. We observe that
real and imaginary parts of amplitudes (\re{eq:M_CSD_def}) indeed probe in
different ways, respectively through cosine and sine Fourier transformations,
twist-2 quark correlation function in a nucleon target at various longitudinal
distances. 

\section{Evolution of coordinate-space skewed parton distributions}
\label{sect:evol_coord_space_dist}

In this section we discuss how the explicit scale dependence of
coordinate-space skewed parton distributions can be derived using arguments
based on the conformal symmetry \cite{Conformal}. Here we consider flavour non-singlet quark
distributions only. A generalization to the flavour-singlet case, including
effects of mixing between quark and gluon operators in the flavor-singlet
sector, will be given in a separate paper.
 
Explicit solutions of evolution equations for momentum-space skewed parton
distributions rely on the fact that, due to the conformal symmetry, one can
find basis of local operators which are multiplicatively renormalized
at the one-loop level. Our aim here is to construct similar expansion in the
coordinate space, but in terms of matrix elements of multiplicatively
renormalizable {\em non-local} string operators. As we shall see, the
corresponding 
coefficient functions are given by Bessel functions $J_\mu(x)$, and the whole
procedure can be understood mathematically as the Neumann series expansion of
coordinate-space distributions.

Recall that a function $f(x)$ can be
expanded in a Neumann series according to \cite{Erdeylyi}
\be\lab{eq:Neumann_series}
f(x) = \sum_{n=0}^\infty (2 \nu + 2 + 4 n ) J_{\nu + 1 + 2n}(x) \int_0^\infty
\frac{d\lambda}{\lambda}\, f(\lambda) J_{\nu + 1 + 2n}(\lambda) \, .
\ee
In particular, one finds that $(\beta \xi)^\nu e^{i \omega
  \beta}$ can be decomposed according to the Sonine's
formula \cite{Erdeylyi}:
\be\lab{eq:Neumann_exp}
(\beta \xi)^\nu e^{i \omega \beta} = 2^\nu \, \Gamma\left[\nu\right] \,
\sum_{n=0}^\infty\, i^n (\nu + n)\, C^\nu_n(\frc{\omega}{\xi})\, 
J_{\nu+n}(\beta \xi) \,
\ee
where $\Gamma$ is the Euler gamma function and $C^\nu_n$ denotes the Gegenbauer
polynomial, respectively. Choosing $\nu
= \frc{3}{2}$, and taking the real part of (\re{eq:Neumann_exp}) one
obtains:
\be\lab{eq:Neuman_cos}
\cos{\left(\omega \beta\right)} = \left(\frac{2}{\beta \xi}\right)^\frac{3}{2}
\, 
\Gamma\left[\frac{3}{2}\right] \, 
\sum_{n=0}^\infty\, (-1)^n (\frc{3}{2} + 2n)\,  
C^\frc{3}{2}_{2 n}(\frc{\omega}{\xi})\, J_{\frc{3}{2}+2n}(\beta \xi) \,
\ee
Inserting this expansion in the definition of the coordinate-space skewed quark
distribution (\re{eq:CSD_to_MSD}) and interchanging summation and
integration 
one obtains:
\be\lab{eq:CSD_s_exp_bess}
{\cal F}^S(\beta,\xi;\mu^2) =
\frac{1}{\sqrt{\pi}}\,\left(\frac{2}{\beta \xi}\right)^\frac{3}{2}\,
\sum_{n=0}^\infty (-1)^n (\frc{3}{2}+2n) J_{\frc{3}{2}+2n} (\beta \xi)
\int_0^1 \!\! du
\,F^S(\omega,\xi;\mu^2) \, C^\frc{3}{2}_{2 n}(\frc{\omega}{\xi}) \, .
\ee
Now, note that a Gegenbauer moment \cite{Rad97}
$$
\xi^{2n} \int_0^1 \!\! d\omega
\,F^S(\omega,\xi;\mu^2) \, C^\frc{3}{2}_{2 n}(\frc{\omega}{\xi})
$$ 
is proportional
to the matrix element of multiplicatively renormalizable local conformal
operator \cite{ER78,Conformal}. Its 
scale dependence is therefore given by  
\be\lab{eq:MR_Op_s_evol}
\xi^{2n} \int_0^1 \!\! d\omega
\,F^S(\omega,\xi;Q^2) \, C^\frc{3}{2}_{2 n}(\frc{\omega}{\xi}) = 
L_{2n+1}\, \xi^{2n} \int_0^1 \!\! d\omega
\,F^S(\omega,\xi;\mu^2) \, C^\frc{3}{2}_{2 n}(\frc{\omega}{\xi})
\ee
where 
\be\lab{eq:L_def}
L_{k} =
\left(\frac{\log{(Q^2/\Lambda^2)}}{\log{(\mu^2/\Lambda^2)}}\right)
^{-\frac{\gamma(k)}{b_0}} \, ,  
\ee
Here, $\Lambda$ is the QCD scale. $\gamma(k)$ is the anomalous dimension, 
$$
\gamma(k) = \frc{4}{3}\,\left(3 + \frac{2}{k(k+1)} - 4 (\Psi(k+1) + \gamma_{\rm
E})\right) \, , 
$$
with $\Psi(x) = \frac{d\, {\rm Log}(\Gamma(x))}{dx}$, $\gamma_E$ being the
Euler constant, and $b_0 = 11 - 2/3 {\rm N_F}$.

As it follows, the scale dependence of the coordinate space distribution  
${\cal F}^S(\beta,\xi;\mu^2)$ is given simply by
\be\lab{eq:CSD_s_scale_dep}
{\cal F}^S(\beta,\xi;Q^2) =\!\!
\frac{1}{\sqrt{\pi}}\,\left(\frac{2}{\beta \xi}\right)^\frac{3}{2}\,
\sum_{n=0}^\infty (-1)^n (\frc{3}{2}+2n) 
L_{2 n + 1}
J_{\frc{3}{2}+2n} (\beta \xi)
\int_0^1 \!\! d\omega
\,F^S(\omega,\xi;\mu^2) \, C^\frc{3}{2}_{2 n}(\frc{\omega}{\xi}) \, .
\ee

In the case of ${\cal F}^A(\beta,\xi;\mu^2)$ one can apply a
decomposition of $\sin{(u \beta)}$ analogous to (\re{eq:Neuman_cos}):
\be\lab{eq:Neuman_sin}
\sin{\left(\omega \beta\right)} = \left(\frac{2}{\beta \xi}\right)^\frac{3}{2}
\, 
\Gamma\left[\frac{3}{2}\right] \, 
\sum_{n=0}^\infty\, (-1)^n (\frc{5}{2} + 2n)\,  
C^\frc{3}{2}_{2 n+1}(\frc{\omega}{\xi})\, J_{\frc{5}{2}+2n}(\beta \xi) \,
\ee
to obtain
\be\lab{eq:CSD_a_exp_bess}
{\cal F}^A(\beta,\xi;\mu^2) =
\frac{1}{\sqrt{\pi}}\,\left(\frac{2}{\beta \xi}\right)^\frac{3}{2}\,
\sum_{n=0}^\infty (-1)^n (\frc{5}{2}+2n) J_{\frc{5}{2}+2n} (\beta \xi)
\int_0^1 \!\! d\omega
\,F^A(\omega,\xi;\mu^2) \, C^\frc{3}{2}_{2 n + 1}(\frc{\omega}{\xi}) \, .
\ee
If ${\cal F}^A(\beta,\xi;\mu^2)$ is associated with a flavor
non-singlet quark distribution,
$$
\xi^{2n+1} \int_0^1 \!\! du
\,F^A(\omega,\xi;\mu^2) \, C^\frc{3}{2}_{2 n + 1}(\frc{\omega}{\xi})
$$
is again multiplicatively renormalizable, but instead of
(\re{eq:MR_Op_s_evol}) we have \cite{Rad97}
\be\lab{eq:MR_Op_a_evol}
\xi^{2n+1} \int_0^1 \!\! d\omega
\,F^A(\omega,\xi;Q^2) \, C^\frc{3}{2}_{2 n + 1}(\frc{\omega}{\xi}) = 
L_{2n+2}\, \xi^{2n+1} \int_0^1 \!\! d\omega
\,F^A(\omega,\xi;\mu^2) \, C^\frc{3}{2}_{2 n + 1}(\frc{\omega}{\xi}) \, . 
\ee
The resulting scale dependence of ${\cal F}^A(\beta,\xi;Q^2)$ is governed
simply by
\be\lab{eq:CSD_a_scale_dep}
{\cal F}^A(\beta,\xi;Q^2) =\!\!
\frac{1}{\sqrt{\pi}}\,\left(\frac{2}{\beta \xi}\right)^\frac{3}{2}\,
\sum_{n=0}^\infty (-1)^n (\frc{5}{2}+2n) 
L_{2 n + 2}
J_{\frc{5}{2}+2n} (\beta \xi)
\int_0^1 \!\! d\omega
\,F^A(\omega,\xi;\mu^2) \, C^\frc{3}{2}_{2 n + 1}(\frc{\omega}{\xi}) \, .
\ee

Now we show that equations (\ref{eq:CSD_s_scale_dep}) and
(\ref{eq:CSD_a_scale_dep}), can be naturally understood as expansions of
coordinate-space skewed quark distributions in terms of matrix elements of
non-local, multiplicatively renormalizable, conformal operators. Indeed,
applying (\re{eq:Neumann_series}) one can rewrite (\ref{eq:CSD_s_scale_dep})
and (\ref{eq:CSD_a_scale_dep}) as a Neumann-type series:
\baa\lab{eq:CSD_scale_dep_nonl}
{\cal F}^S(\beta,\xi;Q^2) &=&\!\! \beta^{-\frac{3}{2}}
\sum_{n=0}^\infty (3+4n) 
J_{\frc{3}{2}+2n} (\beta \xi)
L_{2 n + 1}
\int_0^\infty d\lambda \sqrt{\lambda}
\,{\cal F}^S(\lambda,\xi;\mu^2) \, J_{\frc{3}{2}+2n}(\lambda \xi)
\nonumber \\
{\cal F}^A(\beta,\xi;Q^2) &=&\!\! {\beta^{-\frac{3}{2}}}
\sum_{n=0}^\infty (5+4n) 
J_{\frc{5}{2}+2n} (\beta \xi)
L_{2 n + 2}
\int_0^\infty d\lambda \sqrt{\lambda}
\,{\cal F}^A(\lambda,\xi;\mu^2) \, J_{\frc{5}{2}+2n}(\lambda \xi)
\nonumber \\
\eaa
We are now in position to make relation to conformal symmetry explicit.
Let us start from the obvious identity
\be\lab{eq:identity}
O(\alpha,\beta) = \int_{-\infty}^\infty d\alpha^\prime  \int_0^\infty
d\beta^\prime\, \delta(\alpha-\alpha^\prime) \delta(\beta-\beta^\prime)
O(\alpha^\prime,\beta^\prime) \, . 
\ee
Applying (\re{eq:Neumann_series}) one finds a representation of a
$\delta$-function in terms of a Neumann series
\be
\beta \delta(\beta-\beta^\prime) = \sum_{n=0}^\infty (2\nu+2+4n) 
J_{\nu+1+2n}(\beta)J_{\nu+1+2n}(\beta^\prime) \, .
\ee
Inserting this expansion into (\re{eq:identity}) 
one finds that the string
operator $O(\alpha,\beta)$ can be decomposed as
\be\lab{eq:conf_exp_1}
O(\alpha,\beta) = \beta^{-\frac{3}{2}}
\int_{-\infty}^\infty \frac{dk}{2\pi}\, e^{-ik\alpha}
\sum_{n=0}^\infty (1+2j) 
J_{\frc{1}{2}+j}(|k|\beta) S(\frc{1}{2}+j,k;\mu^2)
\ee 
in terms of conformal string operators $S(\frc{1}{2}+j,k;\mu^2)$. Here we have
introduced a variable $j=\frc{1}{2}+\nu+2n$ which will be identified below
with the conformal spin. 
\noindent Operators
\be\lab{eq:conf_op_S}
S(\frc{1}{2}+j,k;\mu^2) = \int_{-\infty}^{\infty} d\alpha \, e^{i k \alpha}\,
\int_0^\infty d\beta \, \sqrt{\beta} J_{\frc{1}{2}+j} (|k|\beta)
\, O(\alpha,\beta) \, ,
\ee
introduced first in
\cite{BalBra88}, form a representation of a conformal group and are therefore
multiplicatively renormalizable at a one-loop level \cite{BalBra88}, i.e.
\be\lab{eq:conf_op_S_reno}
S(\frc{1}{2}+j,k;Q^2) = L_j \, S(\frc{1}{2}+j,k;\mu^2) \, .
\ee
Because of symmetry properties in the variable $\beta$, one finds that for
charge-conjugation odd and even components of $O(\alpha,\beta)$, see equation
(\re{eq:O_C_decomp}),  $\nu$ has to be
equal to $\frc{1}{2}$ and $\frc{3}{2}$, respectively. Equivalently, one finds
that corresponding expansions run over operators $S(\frc{1}{2}+j,k;\mu^2)$
with the conformal spin
$j$ which assumes values $j = 2 n +1$, respectively $j = 2 n +2$. 
Combining this
observation with (\re{eq:conf_exp_1}) and (\re{eq:conf_op_S_reno}) one
finds 
\baa\lab{eq:conf__exp_2}
O^+(\alpha,\beta;Q^2) &=& \int_{-\infty}^\infty \frac{dk}{2\pi}\, e^{-ik\alpha}
{\beta^{-\frac{3}{2}}} \sum_{n=0}^\infty (3+4n) L_{2n+1} 
J_{\frc{3}{2}+2n}(|k|\beta) S(\frc{3}{2}+2n,k;\mu^2) \, ,
\nonumber \\
O^-(\alpha,\beta;Q^2) &=& \int_{-\infty}^\infty \frac{dk}{2\pi}\, e^{-ik\alpha}
{\beta^{-\frac{3}{2}}} \sum_{n=0}^\infty (5+4n) L_{2n+2} 
J_{\frc{5}{2}+2n}(|k|\beta) S(\frc{5}{2}+2n,k;\mu^2) \, .
\nonumber \\
\eaa
Taking matrix elements of both sides of the above equations one immediately
reproduces equations (\re{eq:CSD_s_scale_dep}) and (\re{eq:CSD_a_scale_dep}).

Finally, let us note that equation (\ref{eq:CSD_scale_dep_nonl}) can be
rewritten in the form 
\be\lab{eq:mathcom1}
{\cal F}^S(\beta,\xi;Q^2) = 
\int_0^\infty d\lambda
{\cal F}^S(\lambda,\xi;\mu^2) 
\sum_{n=0}^\infty L_{2 n + 1} (3+4n) 
\Psi_n(\beta,\xi) {\bar \Psi}_n(\lambda,\xi)
\ee
where $\Psi_n(\beta,\xi) = \beta^{-\frc{3}{2}} J_{2n+1}(\beta \xi)$ and 
${\bar \Psi}_n(\lambda,\xi) = \lambda^{\frc{1}{2}} J_{2n+1}(\lambda \xi)$.
It corresponds to a standard representation
of solution of a non-stationary evolution equation 
\be\lab{eq:evolution}
\frac{\partial}{\partial \tau} {\cal F}^S(\beta,\xi,\tau) = 
\int_0^\infty d\beta'{\hat
H}(\beta,\beta';\xi) {\cal F}^S(\beta',\xi,\tau) \, .
\ee
Here, 'time' $\tau$ is given by
$\frac{1}{b_0}
\log\left(\frac{\log(Q^2/\Lambda^2)}{\log(\mu^2/\Lambda^2)}\right)$. Functions
$\Psi_n(\beta,\xi)$ play a role of 
eigenfunctions of 
the corresponding stationary equation
\be\lab{eq:stat}
{\hat H} \Psi_n = E_n \Psi_n \, .
\ee
Functions 
${\bar \Psi}_n(\beta,\xi)$ are eigenfunctions of the conjugated equation
\be\lab{eq:stat_conj}
{\bar \Psi}_n {\hat H}  = E_n {\bar \Psi}_n \, .
\ee
Energy levels are given by anomalous dimensions $\gamma(2n+1)$. The explicit
form of the hamiltonian ${\hat H}$ can be found in \cite{BalBra88}. Functions
${\bar \Psi}_n$ and $\Psi_m$ are orthogonal with the scalar product given by
the integral 
\be
\langle {\bar \Psi}_n | \Psi_m \rangle = \int_0^\infty d\beta \,
{\bar \Psi}_n(\beta,\xi) \Psi_m(\beta,\xi) = \frac{\delta_{nm}}{3+4n}\, . 
\ee
The sum
\be
R^S(\beta,\beta';\xi) =  
\sum_{n=0}^\infty L_{2 n + 1} (3+4n) 
\Psi_n(\beta,\xi)\, {\bar \Psi}_n(\beta',\xi)
\ee
is the standard representation of the resolvent of the
evolution equation (\ref{eq:evolution}). Evolution of the 
charge-conjugation even quark distribution  ${\cal
F}^A(\beta,\xi;Q^2)$ can be interpreted in an analogous way.

Note that coordinate-space conformal expansions (\ref{eq:CSD_scale_dep_nonl})
and (\ref{eq:conf__exp_2}) have a form which closely resembles a general
solution to the ERBL evolution equation \cite{ER78,BL79}. In momentum
space, however, such an expansion can be written only for distribution
amplitudes i.e., in the case $\xi=1$. In
analogy with the latter, a phenomenological model for coordinate-space
distributions and, ultimately,  physical amplitudes can be obtained by 
choosing expansion coefficients
\baa
a^S_{n}(\xi,\mu^2) &=&
\int_0^\infty d\lambda \sqrt{\lambda}
\,{\cal F}^S(\lambda,\xi;\mu^2) \, J_{\frc{3}{2}+2n}(\lambda \xi) 
\nonumber \\
&=&
\frac{1}{2\sqrt{\pi}} \left(\frac{2}{\xi}\right)^\frac{3}{2} (-1)^n
\int_0^1 \!\! d\omega
\,F^S(\omega,\xi;\mu^2) \, C^\frc{3}{2}_{2 n}(\frc{\omega}{\xi})
\nonumber \\
a^A_{n}(\xi,\mu^2) &=&
\int_0^\infty d\lambda \sqrt{\lambda}
\,{\cal F}^A(\lambda,\xi;\mu^2) \, J_{\frc{5}{2}+2n}(\lambda \xi) 
\nonumber \\
&=&
\frac{1}{2\sqrt{\pi}} \left(\frac{2}{\xi}\right)^\frac{3}{2} (-1)^n
\int_0^1 \!\! d\omega
\,F^A(\omega,\xi;\mu^2) \, C^\frc{3}{2}_{2 n+1}(\frc{\omega}{\xi})
\eaa
according to some phenomenological model of low-energy nucleon structure.
In particular, it would be interesting to try a model in which $a_n$'s are
constructed from linear combinations of their values corresponding to the
forward, $\xi \to 0$, and totally exclusive, $\xi \to 1$, limits. The resulting
momentum-space distributions can be obtained by applying formalism developed in
the next section.

Equations (\ref{eq:CSD_scale_dep_nonl}) and (\ref{eq:conf__exp_2}),
which provide 
a natural solution to the problem of  
LO scale-dependence of coordinate-space skewed quark distributions, belong to
the main results of the present paper. Their form corresponds to an expansion
in terms of a set of orthogonal eigenfunctions which are solutions of the
stationary equations (\ref{eq:stat}) and
(\ref{eq:stat_conj}). Note that a different representation of the same
solution has been
found in \cite{BalBra88}. The latter should be understood as a
representation of the Neumann series by an integral over complex values of
conformal spin $j$.  In the Appendix A we shall
show in details how the Neumann expansion (\ref{eq:CSD_scale_dep_nonl})  can be
obtained starting from the solution found in \cite{BalBra88}. 

Although we intend to discuss numerical aspects of the above algorithm in a
separate 
publication, a short comment on its practical applicability is in
order here. We have
$$
|M^S(\xi)|^2
\ba{l} 
\epsfig{file=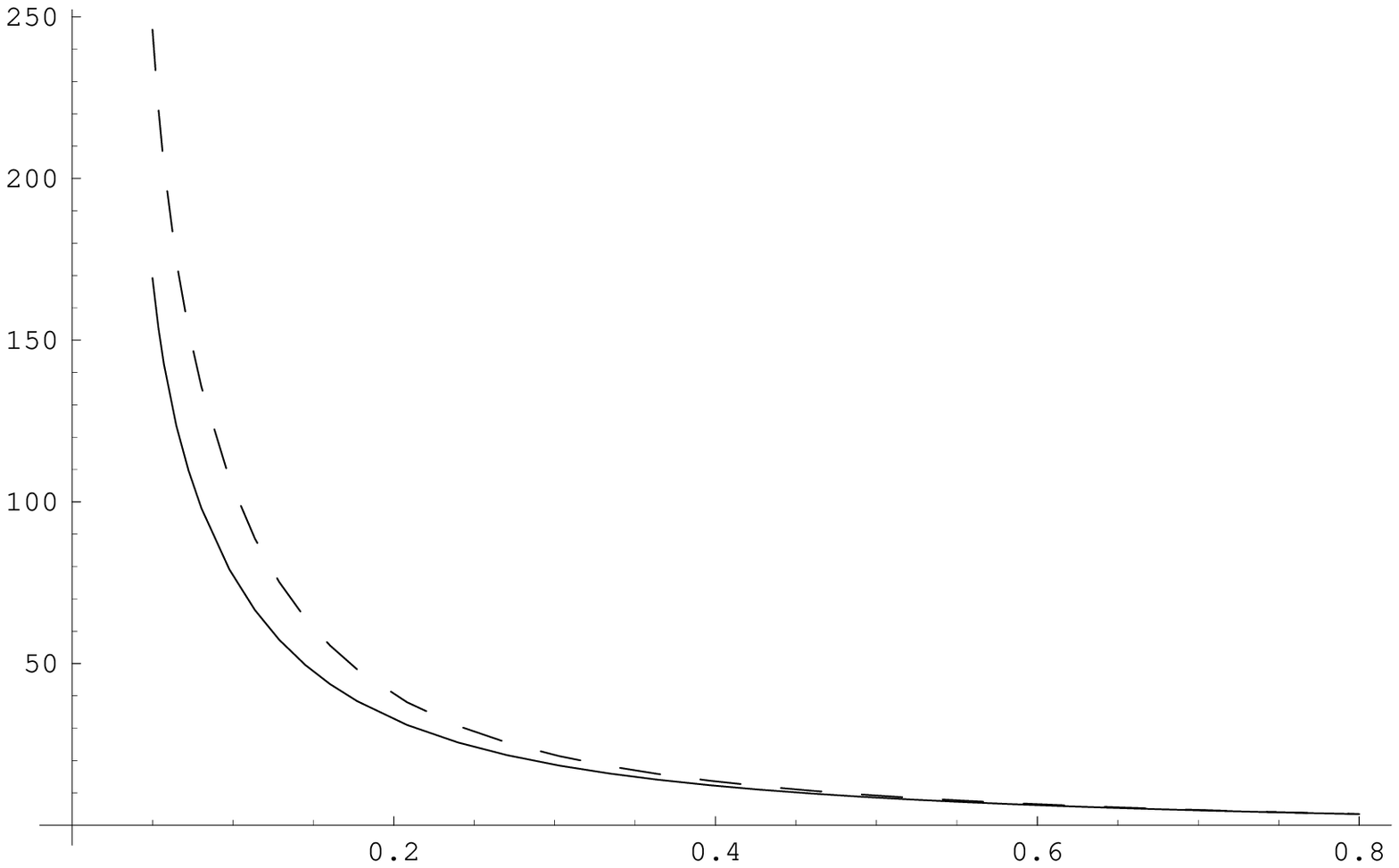,height=90mm,width=100mm}
{}_{{}_{\textstyle\mskip5mu \xi}}
\ea
$$
\vskip1mm
\noindent  
Figure 1. Typical results of evolution of $|M^S(\xi)|^2$ as a function of
$\xi$, 
starting from a $\xi$-independent initial
conditions $F^S(u,\xi;\mu_0^2) = 1.1641 u^{-\frc{1}{2}}(1-u)^{3.5}$. The solid
line denotes $|M^S(\xi)|^2$ at the initial scale  
$\mu_0 = 1.777$ GeV, the dashed line represents $|M^S(\xi)|^2$ evolved to 
$\mu = 10$ GeV.
\vskip3mm \noindent 
checked, using various models of skewed quark
distributions, 
that numerical algorithm for evaluation of physical amplitudes, based on
equations (\ref{eq:CSD_scale_dep_nonl}) and (\ref{eq:M_CSD_def}), gives
accurate and stable
results, see Figure 1 for an example,
except for a case where the variable $\xi$ becomes small. This is 
related to an observation  that a non-zero $\xi$ provides a natural cut-off for
large 
$\beta$ behavior of coordinate-space distributions, which significantly
improves convergence of the Fourier integral (\ref{eq:M_CSD_def}), as
compared to the forward case \cite{ManWeig96}.  

\section {Momentum-space skewed quark distributions}
\label{sect:fourier_transform}

So far, we have discussed skewed parton distributions in coordinate space. In
this section we turn our attention to the momentum-space distributions. Let us
concentrate, for example, on $F^S(u,\xi;Q^2)$. It is related to ${\cal
  F}^S(\beta,\xi;Q^2)$ by the cosine Fourier transformation
(\ref{eq:CSD_to_MSD}). For simplicity of notation, we denote in this section
$F^S(u,\xi;Q^2)$ by $F(u,\xi)$, and the corresponding distribution normalized
at the initial scale $\mu^2$ by $F_0(u,\xi)$.
 
Thanks to the conformal symmetry, the Gegenbauer moments of momentum-space
skewed parton distributions are renormalized multiplicatively, but the
corresponding Gegenbauer polynomials do not form a complete set of functions on
the interval $-1 \le u \le 1$ \cite{Rad97}. 

A solution to the problem of scale dependence of $F(u,\xi)$ based on an
expansion in terms of orthogonal polynomials has been proposed in
\cite{Bel97b,MPW98a} and further developed, up to the NLO, in
\cite{Bel98a}. Recently, another approach has been proposed in
\cite{Shuv99}. In 
the following, we shall discuss an alternative scheme, which originates from
coordinate-space considerations presented in the previous section.

In order to perform the Fourier transformation
let us first split ${\cal F}(\beta,\xi)$ into two pieces,
${\cal F}_{<}(\beta,\xi)$ and ${\cal F}_{>}(\beta,\xi)$:
\be\lab{eq:def_F<_F>}
\ba{l}
\displaystyle
{\cal F}_{<}(\beta,\xi)=
\frac{1}{\sqrt{\pi}}\,\left(\frac2{\beta\xi}\right)^{3/2}\sum_{k=0}^\infty  
L_{2k+1}(-1)^k(3/2+2k)J_{3/2+2k}(\beta\xi)
\int_0^\xi d\omega \, F_0(\omega,\xi)C^{3/2}_{2k}(\frc{\omega}{\xi})\\[7mm]
\displaystyle
{\cal F}_{>}(\beta,\xi)=
\frac{1}{\sqrt{\pi}}\,\left(\frac2{\beta\xi}\right)^{3/2}\sum_{k=0}^\infty
L_{2k+1}(-1)^k(3/2+2k)J_{3/2+2k}(\beta\xi)
\int_\xi^1 d\omega\, F_0(\omega,\xi)C^{3/2}_{2k}({\omega}/{\xi})
\ea
\ee
Obviously, 
\be\lab{eq:def_F<_F>_beta}
{\cal F}(\beta,\xi)={\cal F}_{<}(\beta,\xi)+{\cal F}_{>}(\beta,\xi)\, .
\ee
It can be shown that both series expansions in (\re{eq:def_F<_F>}) are
convergent for all $\xi$ and $\beta$. We denote their Fourier transforms as
$F_{<}(u,\xi)$ and $F_{>}(u,\xi)$, such that 
\be\lab{eq:def_F<_F>_u}
F(u,\xi)=F_{<}(u,\xi)+F_{>}(u,\xi)\, .
\ee
At this stage, splitting of ${\cal F}$ into ${\cal F}_{<}$ and ${\cal F}_{>}$
may seem arbitrary, but we shall demonstrate below that the associated
decomposition of $F$ into $F_{<}$ and $F_{>}$ offers new insights into
the evolution of skewed quark distributions in momentum space.

The next step is to insert the decomposition (\re{eq:def_F<_F>_beta}) into the
cosine Fourier integral (\ref{eq:CSD_to_MSD}). In the case of $F_{<}(u,\xi)$
calculation is simple. Integrating corresponding series in (\ref{eq:def_F<_F>})
term by term one obtains the known result \cite{Bel97a}
\be\lab{eq:F<_series}
\ba{l}
\displaystyle
F_{<}(u,\xi)= \theta(u<\xi)\frac1\xi\left(1-\frac{u^2}{\xi^2}\right)
\sum_{k=0}^\infty\,
L_{2k+1}\,\frac{(3/2+2k)}{(k+1)(1+2k)}C^{3/2}_{2k}({u/\xi})\times
\\[8mm]
\displaystyle
\mskip180mu 
\times \int_0^\xi d\omega\, F_0(\omega,\xi)C^{3/2}_{2k}(\frc{\omega}{\xi}) \, .
\ea
\ee    

The situation is, however, more complicated for ${\cal F}_{>}(u,\xi)$ as
the series 
obtained by performing Fourier
integral  diverges term by term. Instead, one can rewrite $\cos[u\bt]$
as a series expansion in Bessel functions, which
follows from the Sonine's formula (\ref{eq:Neumann_exp}): 
\be\lab{eq:Neumann_Kivel_exp}
\cos[u\beta]= \cos[\beta]+ \sqrt{2\pi\beta}\, (1-u^2)
\sum_{k=0}^\infty\,(-1)^k\frac{(3/2+2k)}{2(k+1)(1+2k)}C^{3/2}_{2k}({u})
J_{3/2+2k}(\beta) \, .
\ee
The detailed derivation of this equation is given in the Appendix.  Now,
inserting 
the above expansion into (\ref{eq:CSD_to_MSD}) and interchanging the
integration with summation one obtains a convergent series:
\be\lab{eq:F>_series}
F_{>}(u,\xi) = 
(1-u^2)\sum_{n=0}^\infty
{\textstyle \frac{(3/2+2n)}{(n+1)(1+2n)} }C^{3/2}_{2n}(u)
\sum_{k= 0}^{n}L_{2k+1}R^n_k(\xi^2)
\int_\xi^1 d\omega\, F_0(\omega,\xi)C^{3/2}_{2k}({w}/{\xi}) \, .
\ee
Here we have introduced the notation
\be\lab{eq:Rnk_def}
R^n_k(\xi^2)=
\frac{(-1)^{(k+n)}\Gamma[3/2+n+k]}{(n-k)!\Gamma[3/2+2k]}\xi^{2k}
{}_2F_1\left[\ba{ll}k-n,\mskip5mu 3/2+n+k\\ \mskip50mu 5/2+2k \ea 
\left|\xi^2\right.\right] \, ,
\ee
with ${}_2F_1$ being the  hypergeometric function. It is easy to see that
$R^n_k(\xi^2)$ 
are polynomials in the variable $\xi^2$ of order $n$ for any $k\le n$. 

As it follows, $F(u,\xi)$ can be split naturally into two pieces, each of
them being represented by a series expansion in Gegenbauer polynomials.
The first part, $F_{<}(u,\xi)$, is defined only in the region $u<\xi$.
It has been noted \cite{Ji97,Rad97,Bel97b} that the evolution of $F_<(u,\xi)$
is 
governed by ERBL-type evolution equation \cite{ER78,BL79}. The form of
$F_<(u,\xi)$ is identical 
with a distribution amplitude $\phi(X)$ with $X = u/\xi$.
The  second part $F_{>}(u,\xi)$ 
can contribute for all $0 \le u \le 1$. 
This part is therefore similar to to the solution based on the expansion of
$F(u,\xi)$ in terms of polynomials orthogonal on a segment $-1 \le u \le
1$ \cite{Bel97b,MPW98a}. 

Formula (\ref{eq:Neumann_Kivel_exp}) can also be used to rewrite
$F_<(u,\xi)$ as an expansion in 
terms of Gegenbauer polynomials $C_{2k}^{\frc{3}{2}}(u)$. In this case, the sum
of $F_<(u,\xi)$ and $F_>(u,\xi)$ reproduces solution found in \cite{Bel97b}. 

Let us now demonstrate that $F(u,\xi)$, given by the sum of $F_<(u,\xi)$ and
$F_>(u,\xi)$, and represented by series expansions (\ref{eq:F<_series}) and
(\ref{eq:F>_series}), respectively, have all properties of the correct
solution. For the sake of clarity we have omitted in the following all details
of underlying calculations. Reader interested in mathematical aspects of this
discussion will find the corresponding material in the Appendix B.

\begin{enumerate}
\item{} {\underline {Expansion of initial conditions}}.\\ 
In the limit $Q^2=\mu^2$, $L_{2k+1}\equiv 1$,
one obtains $F(u,\xi) = F_0(u,\xi)$. In this limit, two pieces in
(\ref{eq:def_F<_F>_u}) give the following contributions:
\baa\lab{eq:Initial_cond}
F_{<}(u,\xi)_{Q^2 = \mu^2} = \theta(u\le\xi)\, F_0(u,\xi)\, ,
\nonumber \\
F_{>}(u,\xi)_{Q^2 = \mu^2} = \theta(u>\xi)\, F_0(u,\xi)\, . 
\eaa
As $F_0(u,\xi)$ is continuous and typically non-zero for $u = \xi$, it
follows immediately that the function $F_<(\xi,\xi)$ is nonzero despite the
factor $\left(1-\frac{u^2}{\xi^2}\right)$ in front of the series in
(\ref{eq:F<_series})! This means that, in general,
the corresponding series expansion in Gegenbauer polynomials 
represents a 
function with the singularity at the end-point $u=\xi$. It belongs therefore
to the functional space $L^p$ with $1<p<2$. Further details related to
the convergence of series expansion in terms of Gegenbauer polynomials can be
found in \cite{Erdeylyi}. 

Note that the presence of the theta-function in (\ref{eq:F<_series}) implies
that partons which initially belonged to the segment $0 \le u \le \xi$ stay
there in the course of the evolution. On the other hand, partons described by
$F_>(u,\xi)$, which belonged initially to the segment $\xi < u \le 1$,
diffuse into the segment $0 \le u \le \xi$. An example of evolution of
$F_>(u,\xi)$ starting from $\xi$-independent initial conditions is shown in
Figure 2. It has been obtained taking 130 terms in the expansion (\ref{eq:F>_series}). 

Moreover,  
considering evolution of
$F_0(u,\xi)$ to infinitesimally larger scale $\mu^2 \rightarrow \mu^2 + \delta
\mu^2$ and interpreting the result as a new initial condition for the next
infinitesimal step, one finds that upon evolution partons cross the border
point $u = \xi$ from right to left side only - once they enter the segment $0
\le u \le \xi$, they never come back. These properties are in agreement with
discussion in \cite{Rad97,Ji97}, based on explicit form of evolution equations.

$$
\ba{l}\mskip-10mu 
\epsfig{file=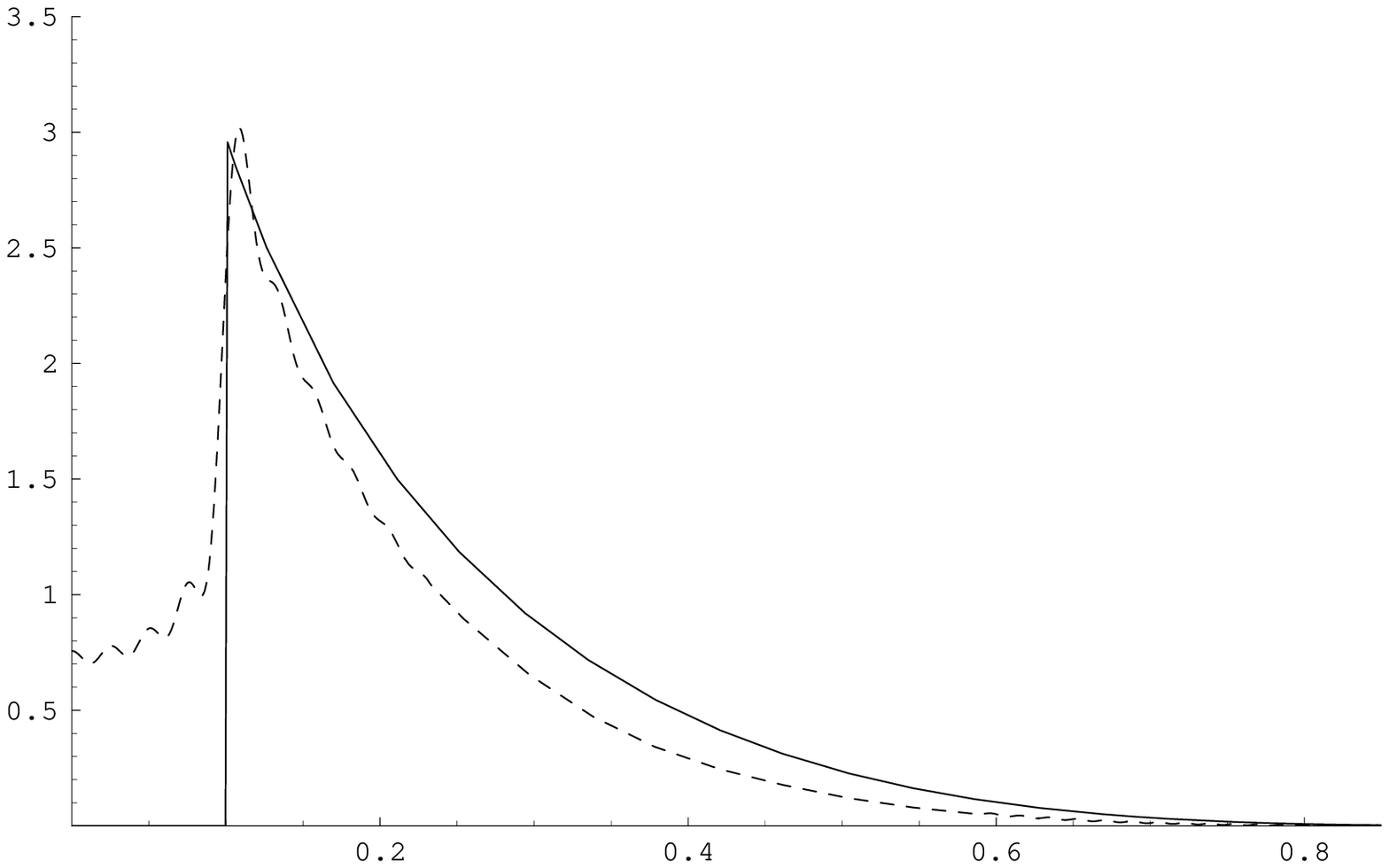,height=55mm,width=75mm} 
{}_{{}_{\textstyle\mskip-10mu \mb{u}}} \mskip10mu
\mskip-5mu
\epsfig{file=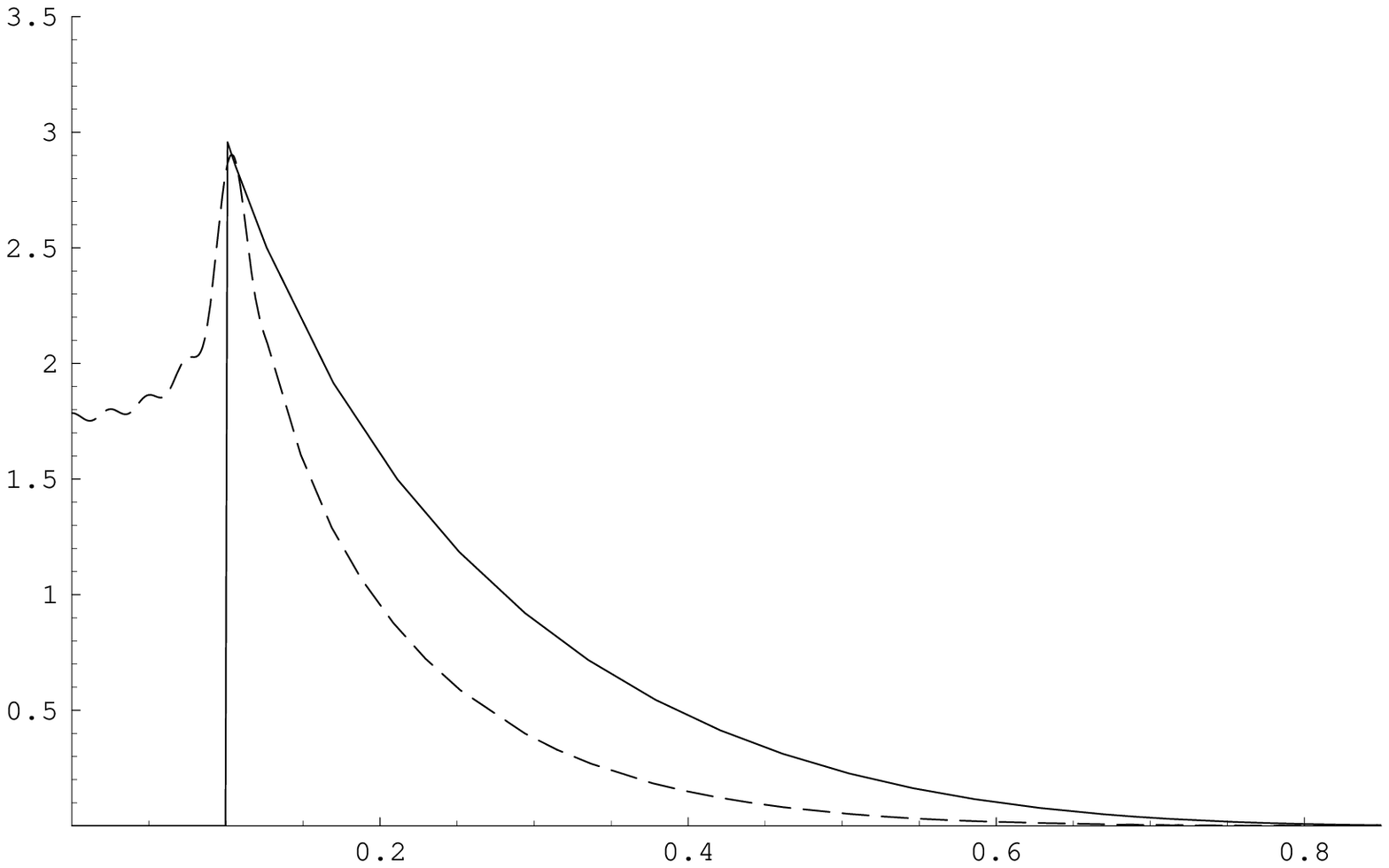,height=55mm,width=75mm}
{}_{{}_{\textstyle\mskip-10mu \mb{u}}}
\\[3mm]
\mskip120mu \mb{Fig. 2a} \mskip260mu \mb{Fig. 2b}
\ea
$$
\vskip1mm 
Figure 2. Evolution of $F_>(u,\xi=0.1)$ starting from a $\xi$-independent
initial conditions 
at the scale $\mu_0 = 1.777$ GeV, solid line. Figures 2a and 2b show
$F_>(u,\xi=0.1)$ evolved to scales $10$, respectively $1000$ GeV, dashed
line. Area under the dashed
curve in the region $u \le \xi$ represents number of partons which diffused
from the region $u > \xi$. As the number of partons which initially occupied
the region $u > \xi$ is conserved, this area is always equal to the area
between dashed and solid lines for $u > \xi$.
\vskip3mm

Finally, both $F_<(u,\xi)$ and $F_>(u,\xi)$ can be treated as two independent
solutions corresponding to boundary conditions (\re{eq:Initial_cond}). Hence,
they should separately obey all constraints imposed by symmetry etc. on a valid
solution. 

\item{} {\underline {Asymptotic limit}}.\\
In the limit $Q^2 \rightarrow \infty$ one finds:
\be
F(u,\xi)\mskip5mu \rightarrow
 \ba{l}{}\\[1mm]
\mskip-40mu \scriptstyle Q^2\rightarrow \infty
\ea 
\theta(u<\xi)\frac{2}{3}\, \frac{1}{\xi}\,
\left(1-\frac{u^2}{\xi^2}\right)
\int_0^1 d\omega \, F_0(\omega,\xi)  \, .
\ee
In particular, the contribution of $F_>(u,\xi)$ reads 
\be
F_>(u,\xi)\mskip5mu \rightarrow
 \ba{l}{}\\[1mm]
\mskip-40mu \scriptstyle Q^2\rightarrow \infty
\ea 
\theta(u<\xi)\frac{2}{3}\, \frac{1}{\xi}\, 
\left(1-\frac{u^2}{\xi^2}\right) 
\int_\xi^1 d\omega \, F_0(\omega,\xi) \, ,
\ee
so in this limit all partons from the segment $\xi < u \le 1$ have diffused
into the segment $0 \le u \le \xi$.

\item{} {\underline {Relation to solutions of ERBL and DGLAP evolution
      equations}}.\\ 
In the limit $\xi\rightarrow 1$ function 
$F(u,\xi=1)$ is identical with the solution of the ERBL evolution equation for
distribution amplitude:
\be
F(u,1)\!\!=\!\! F_{<}(u,1)\!\!=\!\! \left(1-u^2\right)
\sum_{k=0}^\infty\,
L_{2k+1}\,\frac{(3/2+2k)}{(k+1)(1+2k)}C^{3/2}_{2k}(u)\, 
\int_0^1 d\omega\, F_0(\omega,1)C^{3/2}_{2k}(\omega) \, .
\ee

In the limit $\xi\rightarrow 0$ function $F(u,\xi=0)$ is a solution of the 
DGLAP-evolution equation, represented as an expansion in terms of Gegenbauer
polynomials $C^\frc{3}{2}_{2n}(u)$\footnote{This formula have also been
obtained in \cite{Bel97b}}:
\baa
&& F(u,0) = F_>(u,0) =\!\! (1-u^2)\sum_{n=0}^\infty 
\frac{\frc{3}{2}+2n}{(2n+1)(n+1)}C^\frc{3}{2}_{2n}(u) \times
\nonumber \\
&&\times \sum_{k=0}^n L_{2k+1}
\frac{(-1)^{k+n}2^{2k}\Gamma(\frc{3}{2}+n+k)}{(n-k)!(2k)!\Gamma(\frc{3}{2})} 
\int_0^1 d\omega \, \omega^{2k} F_0(\omega,0) \, .
\nonumber \\
\eaa
In particular one finds, as expected, that
\be
\int_0^1 du u^{2k} F(u,0)=L_{2k+1}\int_0^1 du u^{2k} F_0(u,0),\mskip15mu 
 F(u,0)\mskip5mu\rightarrow
\ba{l}{}\\ \scriptstyle\mskip-50mu Q^2\rightarrow\infty \ea
\delta(u) \int_0^1 d\omega \, F_0(\omega,0) \, .
\ee

\item{} {\underline {Conformal constraints}}.\\
As required by conformal symmetry, Gegenbauer moments of the $F(u,\xi)$ are
multiplicatively renormalizable \cite{Rad97,ER78,Conformal}:
\be\lab{eq:mult_ren_geg_mom}
\int_0^1 du\, F(u,\xi)\, C^{3/2}_{2k}(\frc{u}{\xi}) =
L_{2k+1}\int_0^1 du\, F_0(u,\xi)\, C^{3/2}_{2k}(\frc{u}{\xi}) \, .
\ee  
As discussed above, the same is true separately for $F_<(u,\xi)$ and
$F_>(u,\xi)$:
\baa
\int_0^1 du\, F_<(u,\xi)C^{3/2}_{2p}({u/\xi}) &=&
L_{2p+1}\int_0^\xi d\omega\, F_0(\omega,\xi)C^{3/2}_{2p}(\frc{\omega}{\xi})
\nonumber \\
\int_0^1 du\, F_>(u,\xi)C^{3/2}_{2p}({u/\xi}) &=&
L_{2p+1}\int_\xi^1 d\omega\, F_0(\omega,\xi)C^{3/2}_{2p}(\frc{\omega}{\xi}) \,
. 
\eaa
As $L_1 = 1$, in the case of valence quarks the integral from $F(u,\xi)$ over
the whole 
domain is conserved and equal to the number of valence quarks in the target:
\be
\int_0^1 du F(u,\xi)=\int_0^1 du F_0(u,\xi) = N_q \, .
\ee 
In addition, the integrals involving $F_<(u,\xi)$ and $F_>(u,\xi)$ are
separately conserved as well.

\item{} {\underline {Polynomiality}}.\\
Moments of $F(u,\xi)$,
\be\lab{eq:Moments_F}
M^n(\xi) = \int_0^1 du\, u^{2n}\, F(u,\xi)
\ee
can be related to matrix elements of usual, twist-2 local operators of
dimension $3+2n$. Combining this observation with time-reversal invariance and
hermicity one finds that they have to be polynomials in $\xi^2$ of a degree
at most $n$. Note that this is a property of $F(u,\xi)$, but not of
$F_<(u,\xi)$ and $F_>(u,\xi)$ separately. It is preserved by evolution. 
\end{enumerate}

\section {Summary}
The main goal of this paper has been to prepare a theoretical framework for
a quantitative discussion of scale dependence of amplitudes of hard, exclusive
processes with a general hadronic target.
Factorizable amplitudes can
be obtained either as a Mellin convolution, or as a Fourier
integral, of momentum-space, respectively coordinate-space
skewed parton distributions with corresponding Wilson coefficients. 
The explicit  
solution to the problem of the scale-dependence of coordinate-space skewed
parton distributions, obtained in section \ref{sect:evol_coord_space_dist}, has
a structure of a Neumann-series 
expansion of coordinate-space distributions in terms of matrix
elements of non-local, conformal operators. Alternatively, it can be
understood as the
expansion of a QCD string operator in terms of non-local
operators with a definite conformal spin. The particular form of both
expansions explicitly reflect the conformal symmetry of one-loop QCD
evolution equations.  

As discussed in \cite{BalBra88}, the evolution equations for skewed parton
distributions in the coordinate-space
representation have a simple form, analogous to the ERBL evolution
equation for distribution amplitudes. Accordingly,
the solution found here represents a series expansion in
term of a set 
of 
orthogonal eigenfunctions of the evolution kernel in the
coordinate-space. It
is related to the solution in terms of an integral over complex values of
conformal spin $j$, found in \cite{BalBra88}.

Performing Fourier transformation between coordinate- and momentum-space
parton distributions naturally leads to a scheme in which the momentum-space
initial 
conditions are split into two parts, involving partons with
average momentum fractions smaller and larger than the asymmetry parameter. The
former 
never leave their initial domain, while the latter 'diffuse' in the course of
the 
evolution from one domain to the other, speading over the whole interval. We
have used our explicit solution to 
investigate this process numerically and found that the spreading occurs rather
slowly. 

Preliminary numerical studies demonstrate that the coordinate-space approach
leads to stable numerical results for scale-dependence of physical amplitudes
away from the diffractive region.

\bigskip

\noindent
{\bf Acknowledgments:} 

\noindent
N.K. would like to thank the Humboldt Foundation for their financial
support. We gratefully acknowledge discussions with V. Braun, G. Piller,
P. Pobylitsa, 
M.V. Polyakov, A. Radyushkin, W. Weise and C. Weiss. 
\\

\section*{Appendix A.} 

In what follows we will demonstrate how the Neumann expansion
(\ref{eq:CSD_s_scale_dep}) of the charge-conjugation odd skewed quark
distribution follows from the operator solution to one-loop QCD evolution
equations found in \cite{BalBra88}. As in Section
\ref{sect:fourier_transform}, we will denote in the 
following ${\cal F}^S(\bt,\xi;\mu^2)$ simply by ${\cal F}(\bt,\xi)$. The
authors of \cite{BalBra88} argued that 
conformal string operators
\be\lab{eq:conf_string_op_1}
S(\frc{1}{2}+j,k;\mu^2) = \int_{-\infty}^{\infty} d\alpha \, e^{i k \alpha}\,
\int_0^\infty d\beta \, \sqrt{\beta} 
Z_{\frc{1}{2}+j} (|k|\beta) O(\alpha,\beta;\mu^2) \, ,
\ee
where $Z_\nu$ is a solution of Bessel differential equation, form a
representation of the collinear conformal group $SO(2,1)$ and are therefore
renormcovariant. Moreover, applying the Lebedev-Kantorovich transformation
\cite{Lebedev}, they have shown that a flavour non-singlet QCD quark string
operator $O(\alf,\bt;Q^2)$ can be decomposed in terms of
$S({\textstyle\frac{1}{2}} +j,k;\mu^2)$ as:
\be\lab{eq:BalBraun}
O(\alf,\bt;Q^2)=- \beta^{-3/2} \int\frac{dk}{4\pi}e^{-ik\alf}
\int_{\mskip-25mu{}_{{}_{\scriptstyle \delta-i\infty}}}
^{\mskip-25mu{}^{{}^{\scriptstyle \delta+i\infty}}}
\mskip-10mu dj(j+{\textstyle \frac12})
J_{j+ \frac12}(|k|\beta)
L_j
S({\frc{1}{2}} +j,k;\mu^2) \, .
\ee
The Lebedev-Kantorovich transformation requires that
$S(\frac{1}{2}+j,k;\mu^2)$, the multiplicatively
renormalized conformal string operator (\re{eq:conf_string_op_1}), is defined
by taking $Z_\nu = H^{(2)}_\nu$, the Bessel function of the second kind
\cite{Erdeylyi}.  
%
%
The contour of integration over $j$ runs along the imaginary
axis. $\delta$ is a real number chosen in such a way that the contour lies
inside the 
stripe parallel to the imaginary axis, where the integrand is an
analytic function of $j$.   

Matrix elements of the conformal string operator $S(\frc{1}{2}+j,k;\mu^2)$ can
be easily found with the help of (\re{def:corr_F_xi}):
\be\lab{eq:S_matr_elem}
\ba{l}
\displaystyle
\left \langle n(P',S')\right| 
S(\frac{1}{2}+j,k;\mu^2)
\left|n(P,S)\right \rangle =\bar N(P',S')\,\hat z\, N(P,S)\times
\\[4mm] \displaystyle \mskip80mu
\times 2 \pi \ \delta(k - \frac{r}{2}\cdot z)
\int_0^\infty d\beta \, \sqrt{\beta} H^2_{\frac{1}{2}+j} (k\beta)
({\cal F}^S(\beta,\xi;\mu^2) + i \, {\cal F}^A(\beta,\xi;\mu^2)) \, .
\ea
\lab{mat}
\ee
Finally, taking matrix elements of the both sides of the decomposition
(\re{eq:BalBraun}) we arrive at the following representation for the
skewed quark distribution ${\cal F}(\bt,\xi)$:
\be\lab{eq:evolF}
{\cal F}(\bt,\xi;Q^2)=\frac1\pi\bt^{-3/2}
\int_{\mskip-25mu{}_{{}_{\scriptstyle \delta-i\infty}}}
^{\mskip-25mu{}^{{}^{\scriptstyle \delta+i\infty}}}
\mskip-10mu dj(j+{\textstyle \frac12})
J_{j+ \frac12}(\xi\bt)L_{j}
\tilde{S}({\textstyle\frac{1}{2}} +j,\xi;\mu^2) \, ,
\ee
\noindent with
\be\lab{eq:mcso}
\tilde{S}({\frc{1}{2}} +j,\xi;\mu^2)=
\int_0^{\infty}d\bt'\sqrt{\bt'} H^2_{j+\frc{1}{2}}(\xi \beta')
\int_0^1d\om F_0(\om,\xi)\cos[\om\bt'] \, .
\ee
To proceed further, we first integrate by parts over $\omega$. The boundary
terms vanish because $\lim_{u\to 0}\, u F_0(u,\xi) = F_0(1,\xi)=0$. The
resulting integrand has a regular large-$\beta'$ behavior, and therefore it is
possible to interchange integrations over $\omega$ and $\beta'$ to obtain:
\be\lab{eq:mso1}
\ba{l}
\displaystyle
\tilde{S}({\textstyle\frac{1}{2}} +j,\xi;\mu^2)=
\frac1\pi(-1)^{j/2}\left(\frac2\xi\right)^{1/2}\Gamma[1+j/2,1/2-j/2]I_0(j,\xi)+
\\[5mm]\displaystyle \mskip100mu
+\frac{i}{\sqrt{2\pi\xi}}\frac{\Gamma[1/2+j,1/2-j/2,1/2-j]}
{\Gamma[1+j/2]}I_1(j,\xi)+
\\[5mm]\displaystyle \mskip100mu
+\frac{i(-1)^{j+3/2}}{\sqrt{2\pi\xi}}
\frac{\Gamma[1/2+j,1/2-j]}{(1/2+j)\Gamma[1/2-j/2]}I_2(j,\xi) \, .
\ea
\ee
To simplify the above expression we have introduced a special notation for
functions $I_i(j,\xi)$, i = 1,2,3, which are analytic in the region $Rej>0$:
\be\lab{eq:I123}
\ba{l}
\displaystyle
I_0(j,\xi)=\int_0^\xi d\om(\om/\xi) 
F_0'(\om,\xi)
{}_2F_1\left[\ba{ll}1+j/2,\mskip5mu 1/2-j/2\\ \mskip50mu 3/2 \ea 
\left|\frac{\om^2}{\xi^2}\right.\right]
\\[5mm]\displaystyle 
I_1(j,\xi)=\int_\xi^1d\om\left(\frac\om\xi\right)^j
F_0'(\om,\xi)\frac1{\Gamma[1/2-j]}
{}_2F_1\left[\ba{ll}1/2-j/2,\mskip5mu -j/2\\ \mskip50mu 1/2-j \ea 
\left|\frac{\xi^2}{\om^2}\right.\right]
\\[5mm]\displaystyle
I_2(j,\xi)=\int_\xi^1d\om\left(\frac\xi\om\right)^{j+1}
F_0'(\om,\xi)
{}_2F_1\left[\ba{ll}1/+j/2,\mskip5mu 1/2+j/2\\ \mskip50mu 3/2+j \ea 
\left|\frac{\xi^2}{\om^2}\right.\right] \, .
\ea
\ee 
Here $F_0'(\om,\xi)\equiv\frac{d}{d\om}F_0(\om,\xi)$. The integration contour
over $j$ can be wrapped around the 
real positive axis, yielding the sum of residues in poles at $j=1+2k$ and 
$j=1/2+k$ with $k=0,1,2,...$. The series of poles at $j=1+2k$ originates
from the first two terms in the rhs of (\ref{eq:mso1}). The second series of
poles 
at 
$j=1/2+k$ arises 
from the second and third terms in the rhs of (\ref{eq:mso1}). We have found
that 
residues of poles at $j=\frc{1}{2}+k$ cancel exactly each other, so only
poles at $j=1+2k$ give a non-zero contribution to the final result.
The remaining calculations are tedious
but simple. Finally, after integrating by parts over $\omega$
in the opposite direction and rewriting the hypergeometric functions in
terms of the Gegenbauer polynomials, the Neumann series 
(\ref{eq:CSD_s_scale_dep}) arise. From a purely mathematical
point of 
view  this result illustrates a relation between the Lebedev-Kantorovich
transformation and the Neumann series expansion, as discussed in
\cite{Lebedev}.

\section*{Appendix B.} 

Here we shall give the proof of equation (\ref{eq:Neumann_Kivel_exp}) and
provide necessary details of calculations related to discussion in
section \re{sect:fourier_transform}. 

In order to obtain formula (\ref{eq:Neumann_Kivel_exp}) we first expand
$\cos(u\beta)$ in the 
Neumann series using Sonine's formula (\ref{eq:Neumann_exp}) with $\nu=-1/2$
and $\xi = 1$:
\be\lab{eq:cos1}
\ba{l}
\displaystyle
\cos(u\bt)=-\frac{1}{2}(\frac{\beta}{2})^{\frac12}
\Gamma[-{\scriptstyle \frac12}]J_{-\frac12}(\bt)+
(\frac{\beta}{2})\Gamma[-{\scriptstyle\frac12 }]\sum_{k=1}^\infty
(-1)^k(-1/2+2k)C^{-\frac12}_{2k}(u)J_{-\frac12+2k}(\bt)=\\[3mm]
\displaystyle
\mskip80mu 
=\cos(\bt)+\sqrt{2\pi \bt}\sum_{k=0}^\infty
(-1)^{k+1}(3/2+2k)C^{-1/2}_{2(k+1)}(u)J_{3/2+2k}(\bt) \, .
\ea
\ee 
Now, using known properties of the Gegenbauer and Legendre polynomials
\cite{Ryzhik} one can rewrite $C^{-1/2}_{2(k+1)}(u)$ as:
\be\lab{eq:cos2}
\ba{l}
\displaystyle
C^{-1/2}_{2(k+1)}(u)=\frac{(-1)}{2(k+1)}\left[uC^{1/2}_{2k+1}(u)-
C^{1/2}_{2k}(u) \right]=\frac{(-1)}{2(k+1)}\left[uP_{2k+1}(u)-
P_{2k}(u) \right]=\\[4mm] 
\displaystyle \mskip80mu
=\frac{ (1-u^2)}{2(k+1)(2k+1)}\frac{d}{du}P_{2k+1}(u)=
\frac{(1-u^2)}{2(k+1)(2k+1)}\frac{d}{du}C^{1/2}_{2k+1}(u)=
\\[4mm]\displaystyle \mskip80mu
=\frac{1}{2(k+1)(2k+1)}\, (1-u^2) \, C^{3/2}_{2k}(u) \, .
\ea
\ee 
Inserting this formula into (\ref{eq:cos1}) one immediately obtains equation
(\ref{eq:Neumann_Kivel_exp}).

Next, let us discuss properties of matrices $R^p_k(\xi^2)$, introduced in
equation (\ref{eq:Rnk_def}). In our approach they originate from the cosine
Fourier transformation of $F_>(\beta,\xi)$, performed with the help of
the decomposition (\re{eq:Neumann_Kivel_exp}). They can interpreted as
expansion coefficients of 
a Gegenbauer polynomial $C^{\frc{3}{2}}_{2p}(u)$ in terms of Gegenbauer
polynomials $C^{\frc{3}{2}}_{2p}(u/\xi)$ for arbitrary $\xi$, which for
definiteness will be considered here to be greater than $0$:
\be\lab{eq:geg_expansion}
\ba{l}
\displaystyle
C^{\frac32}_{2p}(u)=\sum_{k=0}^p R^p_k(\xi^2)\, 
C^{\frac32}_{2k}(u/\xi),\mskip5mu
\\[4mm]\displaystyle \mskip80mu
R^p_k(\xi^2)= 
\frac{(-1)^{k+p}\Gamma[3/2+p+k]}{(p-k)!\Gamma[3/2+2k]}\xi^{2k}
{}_2F_1\left[\ba{ll}k-p,\mskip5mu 3/2+p+k\\ \mskip50mu 5/2+2k \ea 
\left|\xi^2\right.\right] \, .
\ea
\ee
Note that similar matrices appeared also in
\cite{Bel97b} as an element
of a solution to the evolution equations for momentum-space skewed parton
distributions based on the expansion of the {\it whole} skewed quark
distribution in terms of polynomials orthogonal on the segment $0 \le u \le 1$.
A straightforward way to obtain (\re{eq:geg_expansion}) is to take into account
that 
polynomials $C^{\frc{3}{2}}_{2p}(u/\xi)$ form an orthogonal basis for even
functions of $u$ on the segment $-\xi \le u \le \xi$, with the weight
function $(1-\frc{u^2}{\xi^2})$.
On the other hand, as $C^{\frc{3}{2}}_{2p}(u)$
is a polynomial in $u$ of a degree $2 p$, equation 
$(\re{eq:geg_expansion})$ ensures that left- and right-hand sides are identical
independently whether $u$ is smaller or larger then $\xi$.

Note that the hypergeometric function ${}_2F_1[a,b;c|\xi^2]$ with $a=k-p$,
$b=3/2+p+k$ and $c=5/2+2k$ is itself a  polynomial
in $\xi^2$ of order $p$ as long as $a=k-p\le 0$, i.e. for all $0\le k\le p$. 
In addition, the expansion inverse to (\re{eq:geg_expansion}) is given by the
same matrix function, but evaluated at $1/\xi$, i.e. $R^p_k(\xi^2)$ has the
following remarkable property:
\be\lab{eq:geg_expansion_inverse}
\sum_{k=l}^{p}R^p_k(1/\xi^2)R^k_l(\xi^2)=\delta_{lp} \, .
\lab{f2}
\ee
Unfortunately, we could not find a representation of $R^p_k(\xi^2)$ in terms of
standard orthogonal polynomials. 

Finally, let us concentrate on mathematical aspects which have not been
discussed in details in section
\re{sect:fourier_transform}. 
  
\ul{Asymptotic limit}: here one finds that as $Q^2\rightarrow \infty$, all
$L_{2k+1}\rightarrow 0  
\mb{ for } k>1$ and $L_1=1$. For $F_<(u,\xi)$ only
the first term in the expansion survives, and one obtains:
\be\lab{eq:as1}
F_<(u,\xi)_{Q^2=\infty} =
\theta(u<\xi)\frac{2}{3}\, \frac{1}{\xi}\,
\left(1-\frac{u^2}{\xi^2}\right)
\int_0^\xi d\omega \, F_0(\omega,\xi)  \, .
\ee 
Analogously, for $F_>(u,\xi)$ one obtains:
\be\lab{eq:as2}
\ba{l}
\displaystyle
F_>(u,\xi)_{Q^2=\infty}=
(1-u^2)  \int_\xi^1 d\om \, F_0(\om,\xi) \times
\sum_{n=0}^\infty\frac{(3/2+2n)}{(n+1)(1+2n)}C^{3/2}_{2n}({u}) R^n_0(\xi^2)\, .
\ea
\ee
To prove that the above series indeed equals to
$\theta(u<\xi)\frac{2}{3}\, \frac{1}{\xi}\,
(1-u^2/\xi^2)$ one should determine coefficients $a_p$ of the 
following expansion:
\be\lab{eq:as3}
\theta(u<\xi)\frac3{2}\, \frac{1}{\xi}\, (1-u^2/\xi^2)=
(1-u^2)\sum_{n=0}^\infty a_n \, C^{3/2}_{2n}({u}) \, 
\ee
and compare them with the coefficients of the series in (\re{eq:as2}).
To this end, we multiply both sides by $C^{3/2}_{2p}({u})$ and integrate over
$u$ to obtain:
\be\lab{eq:as4}
a_p=\frac{(3/2+2p)}{(p+1)(2p+1)}\frac2{3}\, \frac{1}{\xi}
\int_0^\xi du\, (1-u^2/\xi^2)C^{3/2}_{2p}({u}) \, . 
\ee
The last integral can be easily calculated in terms of matrix $R^p_k(\xi^2)$:
\be\lab{eq:as5}
\int_0^\xi(1-u^2/\xi^2)C^{3/2}_{2p}({u}){du}/\xi=
\sum_{k=0}^p\, R^p_k(\xi^2) \int_0^\xi \frc{du}{\xi} \,(1-\frc{u^2}{\xi^2})
C^{\frac32}_{2k}(u/\xi)=R^p_0(\xi^2) \, ,
\ee
so one indeed obtains
\be\lab{eq:as6}
a_p=\frac{(3/2+2p)}{(p+1)(2p+1)}\,  R^p_0(\xi^2) \, .
\ee
\\
\ul{Expansion of initial conditions}: when $Q^2 = \mu^2$, one should reproduce 
the initial distribution $F_0(u,\xi)$ as the evolution operator is equal to
unity. 
Consider first $F_<(u,\xi)$. Setting
$L_{2k+1}\equiv 1$ and interchanging summation and integration one 
obtains from (\re{eq:F<_series}): 
\be\lab{eq:unit1}
F_<(u,\xi)_{Q^2=\mu^2}=\theta(u<\xi)\frac{1}{\xi}\int_0^\xi d\omega \,
F_0(\omega,\xi) \left(1-\frac{u^2}{\xi^2}\right)
\sum_{k=0}^\infty\frac{(3/2+2k)}{(k+1)(1+2k)}
C^{3/2}_{2k}({u/\xi})C^{3/2}_{2k}({\omega/\xi}) \, .
\ee
Considering expansion analogous to (\re{eq:as3}) it is easy to see that the
series above represents a $\delta$-function:
\be
\delta(u/\xi-\om/\xi) = \left(1-\frc{u^2}{\xi^2}\right)
\sum_{k=0}^\infty\frac{(3/2+2k)}{(k+1)(1+2k)}
C^{3/2}_{2k}({u/\xi})C^{3/2}_{2k}({\omega/\xi}) \, ,
\lab{unit2}
\ee
and therefore one obtains
\be\lab{eq:unit3}
F_<(u,\xi)_{Q^2=\mu^2}=\int_0^1 d\om \, F_0(\om\xi,\xi)\delta(\om-u/\xi)=
\theta(u<\xi)\, F_0(u,\xi) \, .
\ee
Consider now $F_>(u,\xi)$,as given by (\ref{eq:F>_series}). In the case
$L_{2k+1} \equiv 1$ the internal sum can be evaluated with the help of
(\re{eq:geg_expansion}): 
\be\lab{eq:unit4}
\ba{l}
\displaystyle
F_{>}(u,\xi)_{Q^2=\mu^2}=
(1-u^2)\sum_{n=0}^\infty\frac{(3/2+2n)}{(n+1)(1+2n)}C^{3/2}_{2n}({u})
\int_\xi^1 d\om F_0(\om,\xi)C^{3/2}_{2n}({\om})=
\\[8mm]\displaystyle \mskip85mu
=\int_\xi^1 d\om \, F_0(\om,\xi)\delta(\om-u)=\theta(u>\xi)\, F_0(u,\xi)
\ea
\ee
Combining (\ref{eq:unit3}) and (\ref{eq:unit4}) one immediately finds the
desired answer: 
\be\lab{eq:unit5}
F(u,\xi)_{Q^2=\mu^2}=F_0(u,\xi) \, .
\ee

\ul{Conformal constraints}: here we check explicitly that conformal
moments of $F(u,\xi)$ are multiplicatively renormalized:
\be\lab{eq:mom1} 
\int_0^1 du \, F(u,\xi)C^{3/2}_{2k}(\frc{u}{\xi})=
L_{2k+1}\int_0^1 d\omega\, F_0(\omega,\xi)C^{3/2}_{2k}(\frc{u}{\xi})  
\ee
The case of $F_<(u,\xi)$ is easy. Integrating (\ref{eq:F>_series})
term by term one gets
\be\lab{eq:mom2}
\int_0^1 du\, F_<(u,\xi)C^{3/2}_{2p}({u/\xi})=
L_{2p+1}\int_0^\xi d\omega\, F_0(\omega,\xi)C^{3/2}_{2p}(\frc{\omega}{\xi})
\ee
In case of $F_{>}(u,\xi)$ one can apply (\re{eq:geg_expansion}) and
(\re{eq:geg_expansion_inverse}) to obtain:
\be\lab{eq:mom3}
\ba{l}
\displaystyle
\int_0^1 du F_>(u,\xi)C^{3/2}_{2p}({u/\xi})=
\sum_{l=0}^{p}R^p_l(1/\xi^2)\int_0^1 du F_>(u,\xi)C^{3/2}_{2l}({u})=
\\[6mm]\displaystyle \mskip80mu
=\sum_{n=0}^{p}R^p_n(1/\xi^2)\sum_{k= 0}^{n}L_{2k+1}R^n_k(\xi^2)
\int_\xi^1 dwF_0(w,\xi)C^{3/2}_{2k}({w}/{\xi})=
\\[6mm]\displaystyle \mskip80mu
=\sum_{k= 0}^{p}L_{2k+1}\int_\xi^1 dwF_0(w,\xi)C^{3/2}_{2k}({w}/{\xi})
\sum_{n=k}^{p}R^p_n(1/\xi^2)R^n_k(\xi^2)=
\\[6mm]\displaystyle \mskip80mu
=L_{2p+1}\int_\xi^1 dwF_0(w,\xi)C^{3/2}_{2p}({w}/{\xi}) \, .
\ea
\ee
Sum of (\ref{eq:mom2}) and (\ref{eq:mom3}) gives explicitly  (\ref{eq:mom1}). 

\ul{Polynomiality}: Assume that moments $M_0^n(\xi)$ are
polynomials in $\xi^2$ of a degree $n$. Then, expanding $u^{2n}$ in terms of
$C^{3/2}_{2p}({u}/{\xi})$ and using (\ref{eq:mom1}) one finds 
\be
\int_0^1 du\, u^{2n} F(u,\xi) = \frac{(2n)!}{2^{2n}}
\Gamma\left[\frac{3}{2}\right] \sum_{k=0}^\infty 
\frac{(2n-2k+3/2)}{k!\Gamma[2n-k+5/2]} L_{2n-2k+1}
\xi^{2n}
\int_0^1 d\omega\, F_0(\omega,\xi)C^{3/2}_{2p}(\frc{\omega}{\xi}) \, .
\ee
The last integral can be expressed in terms of $M_0^n(\xi)$ as
\be
\xi^{2n}
\int_0^1 d\omega\, F_0(\omega,\xi)C^{3/2}_{2p}(\frc{\omega}{\xi})=
\frac{1}{\Gamma[\frac{3}{2}]} \sum_{p=0}^{n-k}
\frac{2^{2n-2k-2p}\Gamma[2n-2k-p+3/2]}{p!(2n-2k-2p)!} 
\xi^{2k+2p} M_0^{2n-2k-2p}
\, .
\ee
Collecting all terms, one immediately finds that polynomiality is preserved by
evolution. 

Finally, let us give some arguments in support of the  convergence of the
Gegenbauer series (\ref{eq:F<_series}) and  (\ref{eq:F>_series}). Strictly
speaking, we have shown that expansions are convergent for the case  
$Q^2=\mu^2$, or $L_{2k+1} \equiv 1$ only. We expect, however that evolution
to higher scales will make the convergence only better because of the known 
fact that for large $k$, $L_{2k+1} \sim k^{-c}$ with $c > 0$:
\be
L_{2k+1}\mskip5mu\sim
\ba{l}{}\\[1mm] \scriptstyle\mskip-40mu k\rightarrow\infty \ea
const\cdot k^{-4C_{F}/b_0
\ln\left[\frac{\ln(Q/\Lambda)}{\ln(\mu/\Lambda)}\right]} \, .
\ee  
%


\end{document}